\documentclass[11pt,twoside]{article}

\usepackage{asp2006}
\usepackage{graphics}
\usepackage{epsfig}
\usepackage{lscape}

\def\ltsim{\; \raise0.3ex\hbox{$<$\kern-0.75em \raise-1.1ex\hbox{$\sim$}}\; }
\def\gtsim{\; \raise0.3ex\hbox{$>$\kern-0.75em \raise-1.1ex\hbox{$\sim$}}\; }

\begin{document}
\title{AGN Feedback in groups and clusters of galaxies}   
\author{Somak Raychaudhury$^1$, Simona Giacintucci$^2$, Ewan O'Sullivan$^2$, 
Jan Vrtilek$^2$, 
Judith Croston$^3$,  
Ramana Athreya$^4$, 
Larry David$^2$ \& Tiziana Venturi$^5$}
\affil{$^1$University of Birmingham, UK;
$^2$Harvard-Smithsonian Center for Astrophysics, USA;
$^3$University of Hertfordshire, UK;
$^4$NCRA--TIFR, Pune, India;
$^5$Istituto di Radioastronomia -- INAF, Bologna, Italy}

\begin{abstract} 
The lack of very cool gas at the cores of groups and
clusters of galaxies, even where the cooling time is significantly
shorter than the Hubble time, has been interpreted as evidence of
sources that re-heat the intergalactic medium. Most studies of rich
clusters adopt AGN feedback to be this source of heating. 
From ongoing GMRT projects
involving clusters and groups, we demonstrate how low-frequency GMRT radio
observations, together with Chandra/XMM-Newton X-ray data, present a
unique insight into the nature of feedback, and of the 
energy transfer between the AGN and the IGM.
\end{abstract}


\section{Introduction} 

Simulations of structure formation, using gravitational interactions
between dark matter particles, reproduce, to reasonable accuracy, the
large-scale properties of the observed web of galaxies and clusters
(e.g. Springel et al. 2005). However, when gas physics are added, the
simulated galaxies fail to match some of the basic observed properties 
(e.g., Croton et al. 2005).  To produce the observed luminosity function
of galaxies, the rate of gas cooling has to be set too high, leading
to a severe overestimation of the fraction of baryons in stars, and an
underestimate of the fraction in the X-ray emitting gas. Thus,
additional sources of heating are required:
gravitational heating, supernovae, mergers and conduction are among plausible
candidates, but most studies of rich clusters adopt AGN heating to be
predominant (see McNamara \& Nulsen 2007 and references therein).

The nature of this feedback, vital to our understanding of galaxy and
structure evolution, is one of the most important unresolved questions
in extragalactic astronomy. In the cores of groups and clusters of
galaxies, radiative cooling times are shorter than the Hubble time
(see Fig.~1a). Thus, {\it cooling flows} should result, yet the
spectral signature of gas cooling to low temperatures is not seen
(Peterson \& Fabian 2006). This should be linked to the same
source of feedback that is required by the simulations.  Furthermore, the
raised entropy of the gas within galaxy groups, compared to that
expected from scaling cluster properties (``symmetry breaking''), shows
evidence of the heating of baryons at modest redshifts before they are
assembled into clusters, providing more support for feedback on
the scale of galaxies (Ponman, Cannon \& Navarro 1999).

\section{Observational study of AGN feedback}

Most studies of feedback have hitherto focused on AGN jet/cavity
systems in the most massive, X-ray luminous galaxy clusters, where the
disturbed structures are most clearly seen. However, most galaxies and
most of the baryonic matter in the Universe reside in substantially
smaller groups, where feedback in principle has the greatest impact on
galaxy formation and evolution. It is vital to study feedback in a
sample with a wide variety of groups {\it and} clusters, if we are to
understand how feedback has influenced the thermal history of galaxies
and the intergalactic medium (IGM), and thus of most of the baryons in
the Universe.

 \begin{figure}
\centering
\includegraphics[width=5.0cm]{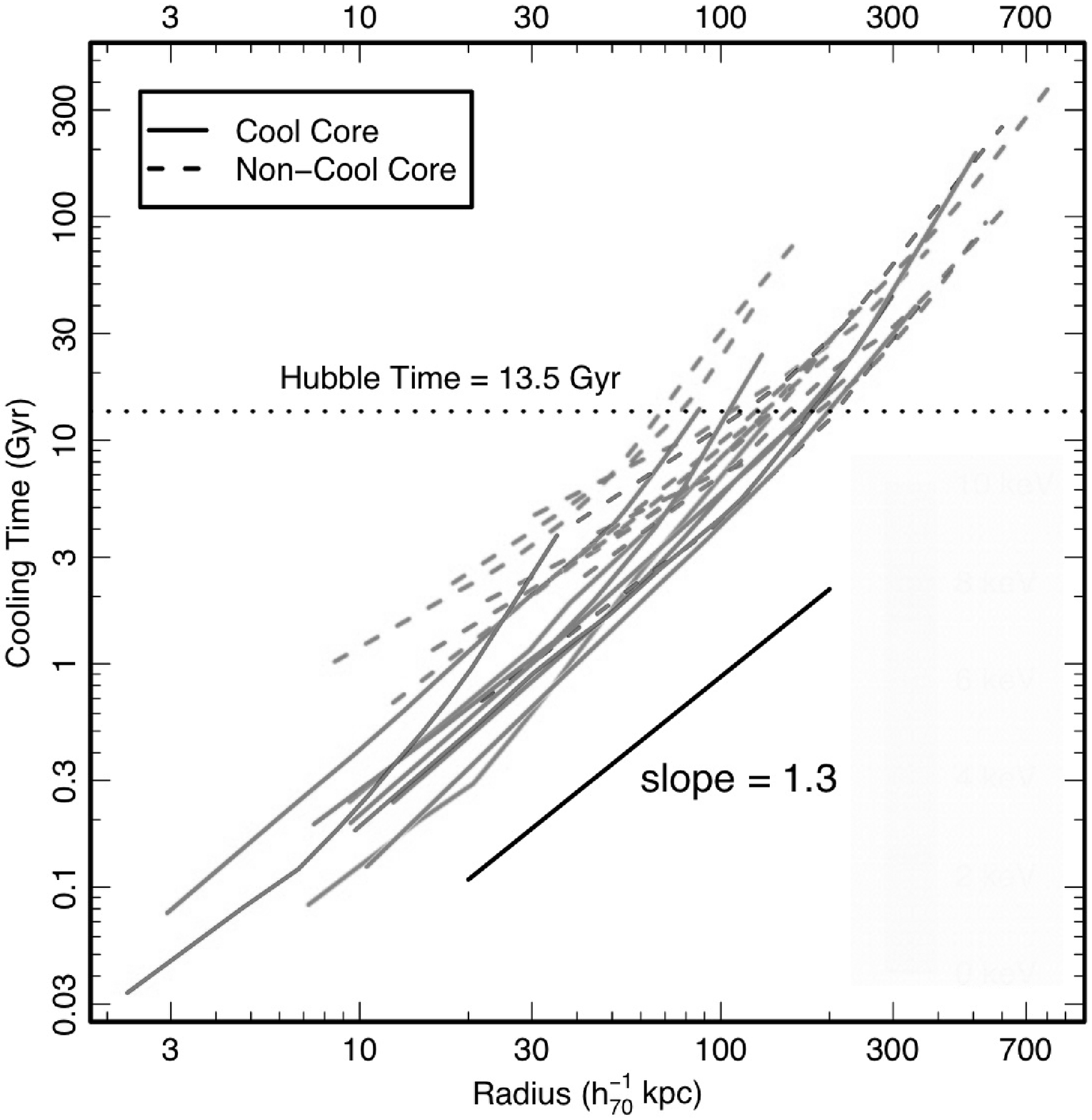} 
\includegraphics[width=5.9cm]{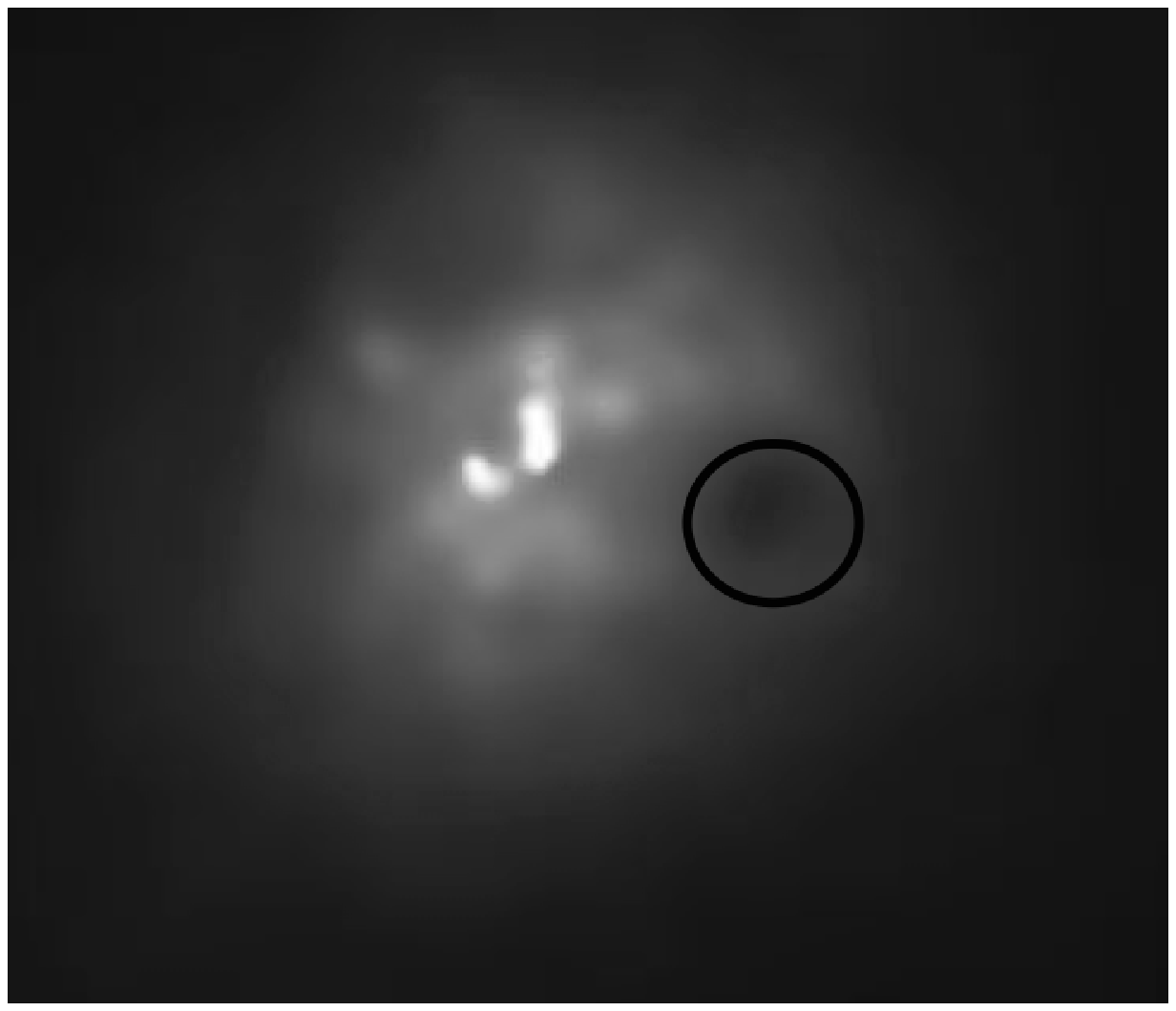}
 \caption{{\itshape (a, left)\/} The cooling time of the
X-ray emitting intergalactic medium at the cores of galaxy clusters
is less than the Hubble time. The dashed lines represent
clusters in which the expected cool core is not found (from
Sanderson et al. 2006). This is true of X-ray emitting
groups of galaxies as well.
 {\itshape (b, right)\/} Cavities, created in the IGM
as a result of the  plasma injected by the central radio source,
are confined by the pressure of the IGM. They buoyantly
rise through the IGM, 
transferring a significant part of their enthalpy.
Here, in the Chandra X-ray image of the cluster Abell 2597
(McNamara et al. 2001), one such detached (``ghost'') cavity
is marked.
}
 \end{figure}

\subsection{The synergy of X-ray and radio observations}

The IGM in groups and clusters emits X-rays, while the relativistic
particles from the AGN produce radio emission through synchrotron
radiation.  However, no simple relation has been established between
radio power and the energy required to destroy cool cores.  A unified
study of observations in multiple wavebands is required to examine the
detailed physics of this process of transfer of energy between the AGN
and the IGM.  The high resolution of the Chandra observatory reveals
highly disturbed structures in the cores of many clusters, including
shocks, cavities and sharp density discontinuities.  Comparing with
radio maps, it is clear that AGN jets are associated with many of
these disturbances.  

The most typical configuration is for jets from
the central dominant elliptical of a group or cluster to extend
outwards and inflate lobes of radio-emitting plasma, pushing aside the
X-ray emitting gas of the cluster halo to create cavities in the hot
IGM (Fig.~1b). The enthalpy of a cavity filled with relativistic
plasma is $4pV$, where $p$ and $V$ are the pressure and volume
respectively (McNamara \& Nulsen 2007), and a large fraction of this
can be transferred to the IGM as the buoyant cavities travel outwards
(see, e.g., McNamara \& Nulsen 2007, Jetha et
al. 2008).

 \begin{figure}
\centering
\includegraphics[width=6.0cm]{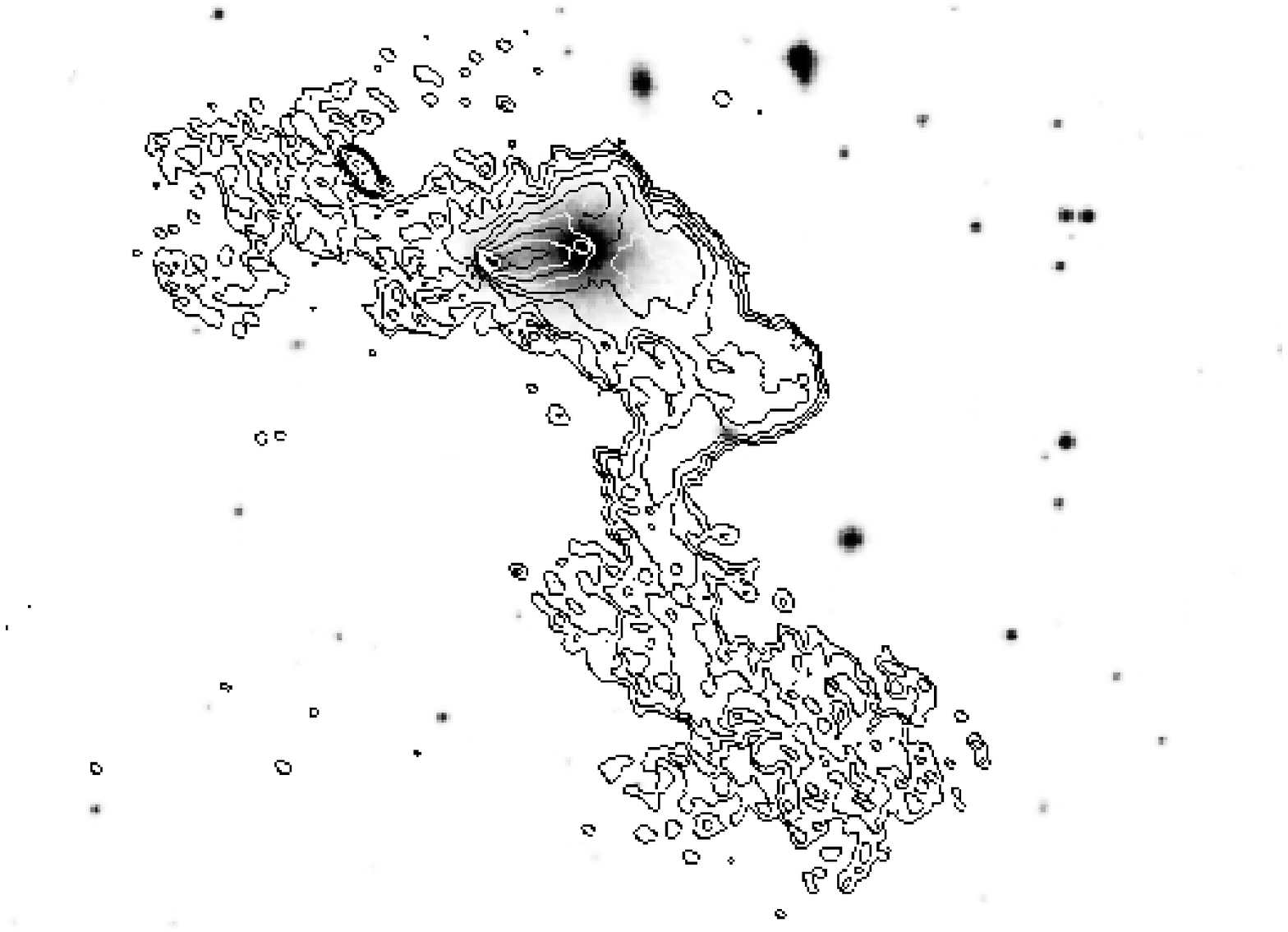}
\includegraphics[width=6.0cm]{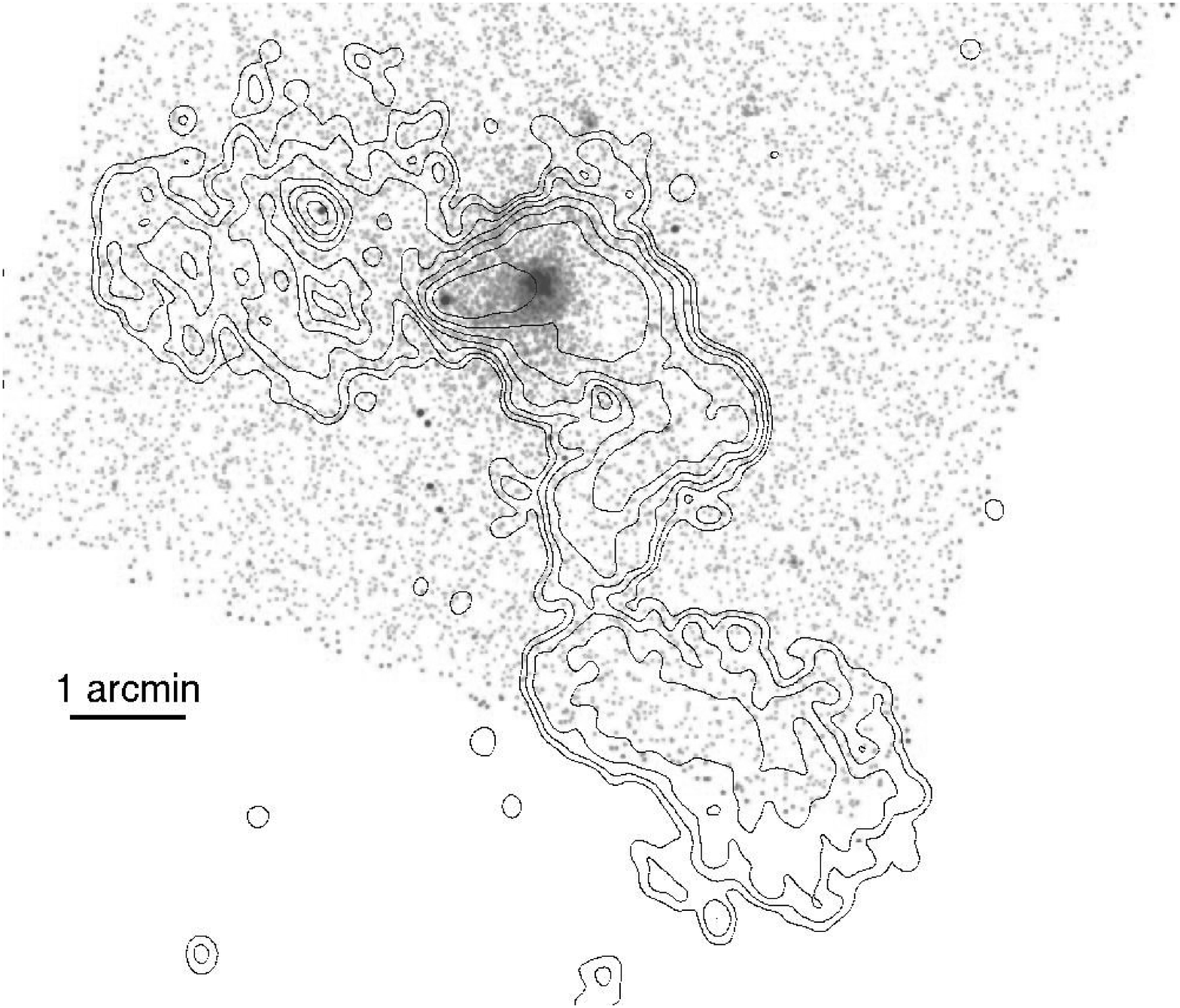}
 \caption{
In the NGC 741 fossil group, the jets are emitted from
a non-central dwarf galaxy, NGC 742. 
{\itshape (a, left)\/} Our 610~MHz GMRT contours, superposed
on an optical image, and
{\itshape (b, right)\/} Our 235~MHz GMRT contours, superposed
on the 0.5-2~keV Chandra X-ray image, show far more
extended radio emission than previous VLA maps. Lowest contours are at
0.15 and 0.9 mJy/beam respectively.} 
 \end{figure}

\subsection{GMRT studies of galaxy groups}
\label{Obsgrp}

Groups have crucial advantages over richer clusters for studies of
feedback -- the gas in their relatively shallow potential wells is
expected to respond more readily to energy input, and the lower
temperature of the IGM would lead to stronger metal emission lines.
Our in-depth study of 18 elliptical-dominated galaxy groups, which
show structures indicative of AGN interaction in deep Chandra/XMM
X-ray images, at 3 GMRT frequencies (235-610 MHz), demonstrates
(Vrtilek et al. 2009, O'Sullivan et al. 2009) that by combining deep
X-ray and GMRT observations, the extent of the AGN outbursts can be
better ascertained, and the chance of detecting earlier outbursts
enhanced, than in $>$1~GHz observations (Giacintucci et al. 2008).
These observations (Figs.~2 \& 3a) allow one to measure the spectral
index as a function of position along the jets, and thus to carefully
model the extended emission, and, using equipartition values for the
ambient magnetic field, to derive the age of the outburst.  This leads
to an estimate of the duty cycle of AGN activity, and helps to
decouple multiple outbursts, if present.  From all of this, in turn,
an accurate estimate of the energy of the AGN outburst can be made.

\subsection{GMRT studies of feedback in galaxy clusters}
\label{Obsclus}
Previous studies of feedback in clusters (e.g. Birzan et al. 2004,
Dunn et al. 2005, Fabian et al. 2006, McNamara \& Nulsen 2007) have
concentrated on the core regions of clusters where X-ray images reveal
the presence of cavities and related features.  To assess how
different kinds of radio outbursts contribute to cluster heating, and
how common it is to find AGN activity related to their brightest
galaxies, a statistical sample needs to be studied. We have obtained
GMRT 610~MHz images of all 26 GMRT-accessible clusters ($z\!<\! 0.2$)
of the REXCESS sample (B\"ohringer et al. 2007). Data reduction is in
progress: we show a very preliminary map of one of these clusters in
Fig.~3b.  These clusters have extensive multiwavelength observations
(including deep X-ray observations), which we will be able to use in
conjunction with the radio observations to model the behaviour of the
AGN and its effect on the cluster IGM.

 \begin{figure}
\centering
\includegraphics[width=5.5cm]{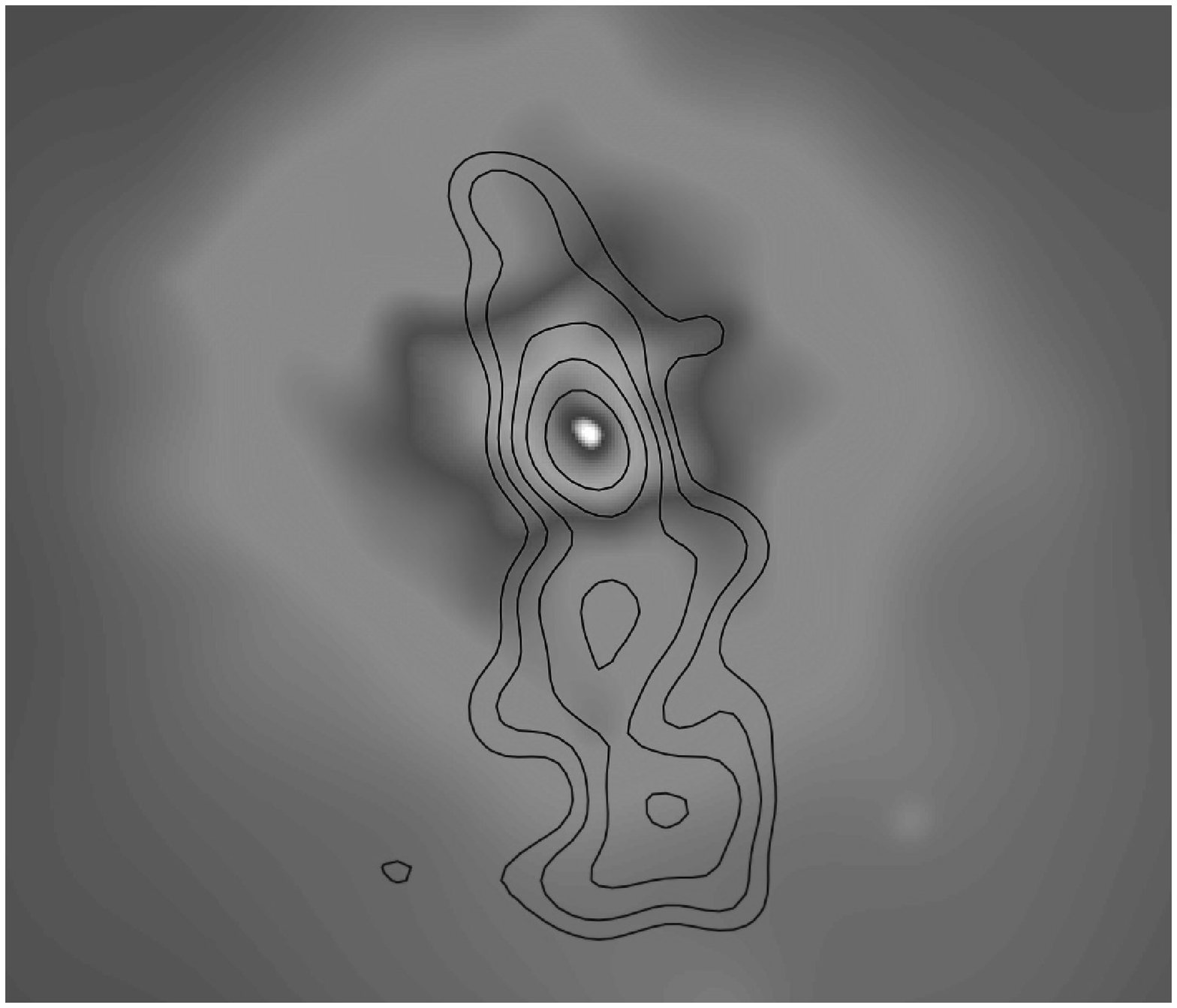}
\includegraphics[width=6.8cm]{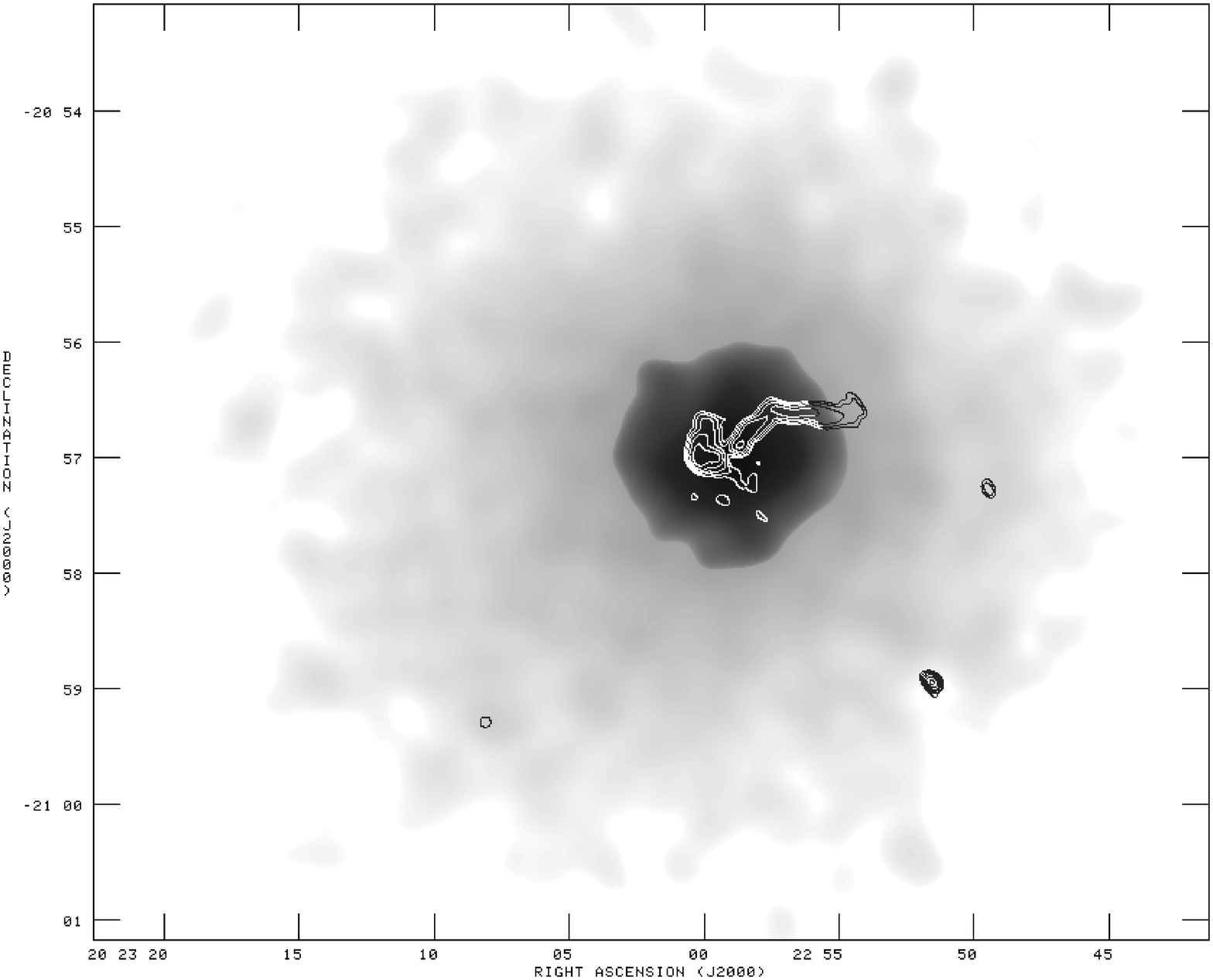}
\caption{
{\itshape (a, left)\/}
Smoothed Chandra X-ray image of the HCG~62 compact group
of galaxies, in which two
  $\sim$8~kpc cavities are detected.  Our GMRT observations (here
  235~MHz contours on the Chandra 0.3--2 keV image) show far more
  extended emission than previous VLA observations (Vrtilek et al. 2002), 
   and evidence of
  progressive spectral ageing along the jets.
{\itshape (b, right)\/} (Preliminary) GMRT 610~MHz contours superposed on 
the XMM-Newton image of the 2.5~keV 
REXCESS cluster RXJ2023.0-2056 at redshift $z=0.056$, 
showing clear detection of jets
associated with the central galaxy.
}
 \end{figure}

\section{Conclusions}   
The most likely sources of feedback in galaxy groups and clusters are
active galactic nuclei. 
However, the radio power of AGN in central galaxies measured at
high radio frequencies is not sufficient to explain the evidence of
heating in the IGM of most groups and clusters. 
Low frequency GMRT observations, along with high-resolution
Chandra X-ray observations, are playing a crucial role in
the detailed study of the history of AGN outbursts
and the mechanism of energy transfer, in an ongoing series of
investigations.

\acknowledgements 
We thank the staff of the GMRT (run by NCRA-TIFR) 
for making these observations possible. 


\end{document}